\documentclass[11pt]{article}

\usepackage[english]{babel}

\usepackage{amsmath,amsthm,amsfonts,amscd,amssymb,eucal,latexsym,slashed}
\usepackage{cite}
\usepackage{epsfig}

\newtheorem{teorem}{Theorem}[section]

\newtheorem{proposition}[teorem]{Proposition}

\newcommand{\RR}{\mathbb R}
\newcommand{\NN}{\mathbb N}

\newcommand{\cA}{{\cal A}}
\newcommand{\cB}{{\cal B}}
\newcommand{\cH}{{\cal H}}
\newcommand{\cL}{{\cal L}}

\newcommand{\cR}{{\cal R}}

\newcommand{\cU}{{\cal U}}

\title{New Light on Infrared Problems:\\[-2mm]  
Sectors, Statistics, Spectrum and All That\footnote{Talk given at 
XVIIth International Congress on Mathematical Physics, 
Aalborg, 6-11 August 2012}} 
\author{Detlev  Buchholz \\[2mm]
Institut f\"ur Theoretische Physik, 
Universit\"at G\"ottingen, \\
37073 G\"ottingen, Germany} 

\date{August 22, 2012}

\begin{document}
\maketitle 

\begin{abstract} \noindent
Within the general setting of algebraic quantum field theory, a new approach
to the analysis of the physical state space of a theory is
presented; it covers theories with long range forces, such as 
quantum electrodynamics. Making use of the notion of charge class,
which generalizes the concept of superselection sector, 
infrared problems are avoided. In fact, on this basis one can determine
and classify in a systematic manner the proper charge content
of a theory, the statistics of the corresponding states and their
spectral properties. A key ingredient 
in this approach is the fact that 
in real experiments the arrow of time gives rise to a 
Lorentz invariant infrared cutoff of a purely geometric nature.
\end{abstract}



\section{Outline} \label{sec1}

The understanding of the sector structure in quantum field theories
with long range forces, such as quantum electrodynamics, is a longstanding 
problem. Its various aspects have received considerable attention 
in the past, cf.\ for example 
\cite{FrMoSt,Bu1,St,Ha} and references quoted there. 
But a fully satisfactory solution has
not yet been accomplished. We report here on a novel approach 
which has been developed in collaborations with 
Sergio~Doplicher and John~E.~Roberts. It sheds new and            
promising light on this problem. 
 
Recall that a superselection sector is a subspace of the 
Hilbert space of all states of finite energy on which the 
local observables of the theory act irreducibly; hence all 
superselected charges have sharp values and the superposition
principle holds unrestrictedly in each sector. The presence of long range
forces leads to an abundance of different sectors due to 
the multifarious formation of clouds of low energy massless      
particles; in fact their comprehensive analysis and classification 
is unfeasible. In the treatment of models this problem
is frequently circumvented by some \textit{ad hoc} selection of sectors,
\textit{e.g.} by choosing a specific physical gauge, and by an 
inclusive treatment of undetected low energy massless particles. 
This strategy is a meaningful workaround but it is clearly
not a satisfactory conceptual solution of the problem.

In contrast, the sector structure is fully understood in   
theories describing only massive particles
\cite{DoHaRo,BuFr,DoRo}. In these theories the sectors are
in one-to-one correspondence to the dual of some compact 
group which is interpreted as global gauge group. Each sector has 
definite (para--Bose or para--Fermi) statistics and there always  
exist (cone) localized Bose and Fermi fields, respectively,  
which transform as tensors 
under the action of the gauge group and create the sectors from
the vacuum state. Moreover, in spite of the possible non--locality of  
these fields, the spin--statistics theorem and the existence of
collision states have been established in these theories.     
The arguments underlying these results fail, however, in the 
presence of long range forces \cite{Bu1}. 

Our resolution of this problem 
is based on the insight that the
arrow of time enters in a fundamental way in the 
interpretation of the microscopic theory: since it is impossible to
perform measurements in the past, the notion of
superselection sector becomes physically meaningless in the 
presence of massless particles. As will be explained, it has to
be replaced by the notion of charge class which is based on 
a natural equivalence relation between sectors \cite{Bu1}. 
We will focus attention here on the family of 
simple charge classes which covers the electric charge.
It can be shown that there is a physically meaningful way 
to assign to each member of this family definite (Bose or Fermi) 
statistics, there always exists a
corresponding conjugate charge class with 
the same statistics and the family 
of all simple charge classes determines a compact abelian gauge group.
Moreover, there is a natural action of the Poincar\'e group 
on each charge class which is implemented by a 
unitary representation
satisfying the relativistic spectrum condition. The generators
of the time translations, however, do not have the familiar 
interpretation                                                  
as global energy. In fact, they resemble the Liouvillians in 
Quantum Statistical Mechanics since they   
subsume in a gross manner energetic fluctuations of the 
infrared background. 

\section{Input}

The present approach applies to theories of local observables fitting 
into the general algebraic framework of quantum field theory \cite{Ha}. 
Any such theory provides an assignment (net) $\cR \mapsto {\cA}(\cR)$ 
mapping spacetime regions $\cR \subset \RR^4$ to 
unital C*--algebras ${\cA}(\cR)$ generated by the observables localized in 
the respective regions. The C*--inductive limit of these local 
algebras is denoted by $\cA$ and assumed to act on 
the defining vacuum Hilbert space $\cH$ of the theory.
The net satisfies 
the condition of locality (spacelike commutativity)
$$[{\cA}({\cR}_1), \, {\cA}({\cR}_2)] = 0 
\ \ \mbox{if} \ \ {\cR}_1 \mbox{\Large $\times$} {\cR}_2 $$ 
and of covariant automorphic action  $\alpha :           
{\cal P}_+^\uparrow  \rightarrow \mbox{Aut} \, \cA$ 
of the Poincar\'e group  ${\cal P}_+^\uparrow =
\RR^4 \rtimes {\cal L}^\uparrow_+$,  
$$\alpha_{\lambda} \, {\cA}({\cR}) 
= {\cA}({\lambda \cR}) \, , \quad \lambda \in {\cal P}_+^\uparrow \, . $$
There is an, up to a phase unique, vector state $\Omega \in \cH$ 
describing the vacuum and fixing a continuous unitary representation 
$U : {\cal P}_+^\uparrow  \longrightarrow \ \cU(\cH)$ 
through the  relation 
$$ U(\lambda) \, A \, \Omega = \alpha_{\lambda}(A) \,
\Omega,  \quad  \lambda \in  {\cal P}_+^\uparrow, \ A \in \cA \, . $$
The subgroup of spacetime translations satisfies the    
spectrum condition, $\mbox{sp}\,U \upharpoonright 
{\mathbb R}^4 \subset \overline{V}_+$. Thinking primarily of
theories describing interactions of electromagnetic type, we assume
that there exist massless single particle states in $\cH$ (photons)
and there is some mass threshold above which there appear pairs of
massive particles carrying opposite charges (electron--positron pairs 
{\it etc}),
cf.\ Fig.~1. The scattering states of photons are assumed to 
be asymptotically complete below this threshold \cite{Bu1}.

\vspace*{5mm}
\begin{figure}[h] 
\centering
\epsfig{file=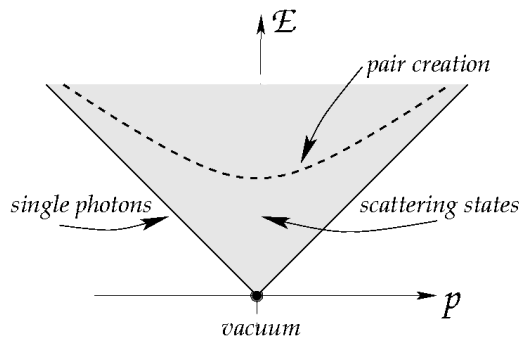,width=70mm}
\caption{Energy--momentum spectrum in vacuum sector} 
\label{fig1} 
\end{figure}

Besides the states in the vacuum sector there exist other elementary 
states of physical interest,  describing charged particles, 
atoms, ions, molecules, \textit{etc}. We adopt here the convenient
point of view that these states are described by 
suitable representations $\pi: \cA \rightarrow \cB(\cH)$ acting all  
on  the same (separable) Hilbert space $\cH$. The (identical)
vacuum representation is denoted by $\iota$. 

\vspace*{2mm} 
\noindent \textbf{Selection criterion for states of interest:} 
The states of elementary systems 
are described by irreducible representations  
$(\pi, \cH)$ of $\cA$ for which there is 
a continuous unitary representation 
$U_\pi: {\mathbb R}^4 \rightarrow {\cal U(H)}$ such that  
$\mbox{Ad} \, U_\pi(x) \circ \pi = \pi \circ \alpha_{x}$ (covariance)
and $\mbox{sp}\,U_\pi \subset \overline{V}_+$
(stability).

\vspace*{2mm} 
Note that it is not required in this criterion 
that the Lorentz transformations are
also implemented. Because this would exclude 
from the outset states carrying an electric
charge \cite{Bu2}. On the other hand there is an                       
abundance of disjoint representations satisfying this criterion 
which, however, differ only by unobservable 
infrared properties as will be explained next.

\section{Charge classes}

Realistic experiments are performed in finite spacetime regions.
Beginning at some appropriate spacetime point one 
performs preparations of states and measurements until sufficient data 
are taken. In principle, subsequent 
generations of experimentalists could continue 
the experiment into the distant future. Thus the maximal regions
where data can be taken are future directed 
lightcones $V$ with arbitrary apex $(t_0, \! x_0)$, 
cf.\ Fig.~2. On the other hand it is impossible to make up for missed     
operations in the past of the initial point $(t_0, \! x_0)$. 
Hence, as a consequence of the ``arrow of time'',
it suffices for the comparison of theory and experiment   
to consider the restrictions of global states to the 
algebra $\cA(V)$ of observables localized in any given 
future directed lightcone~$V$. These restrictions are called partial
states.                          

\vspace*{5mm}
\begin{figure}[h]
\centering
\epsfig{file=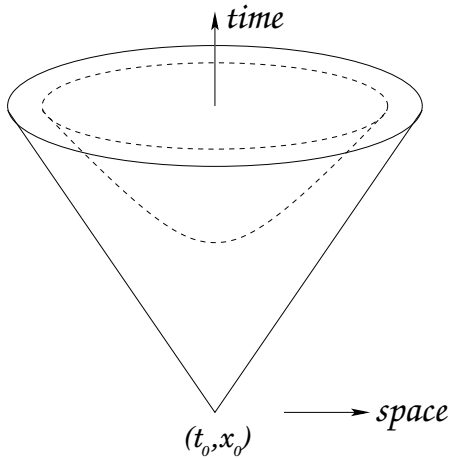,width=60mm}
\caption{Spacetime region $V$ foliated with a  hyperboloid (time shell)} 
\label{fig2}
\epsfxsize80mm
\end{figure}

In massive theories the algebras $\cA(V)$ are irreducible 
\cite{SaWo} and complete information about the 
underlying global states can be 
obtained in any lightcone $V$. The situation is markedly 
different, however, in the  presence of massless 
particles \cite{Bu1}. This is so because   
outgoing radiation created in the past of $(t_0, \! x_0)$ has no observational
effects in $V$ in accordance with Huygens'
principle. As a consequence, infrared clouds cannot sharply be 
discriminated by measurements in any given lightcone  $V$
and the algebras $\cA(V)$ are highly reducible; in fact, their
weak closures $\cA(V)^-$ are factors of type III${}_1$. Whereas 
the infrared sectors cannot be distinguished in any  
lightcone $V$, their total charge can be determined there. This
follows from the fact that 
the charge is tied to massive particles which eventually enter
$V$, unless they are annihilated in 
pairs carrying opposite charges. 
These heuristic considerations suggest to introduce the 
following equivalence relation \cite{Bu1}. \\[2mm] 
\textbf{Definition of charge classes:} \ Let $ (\pi_1, \cH)$, $(\pi_2, \cH)$ 
be representations satisfying the above selection criterion
and let $V$ be any lightcone. The
representations belong to the same charge class if their 
restrictions to $\cA(V)$ are unitarily equivalent, 
$$ \pi_1 \upharpoonright \cA(V) \simeq \pi_2 \upharpoonright \cA(V)
\, .$$

It can be shown that the charge classes do not depend on 
the choice of $V$; moreover, the restricted representations
are primary, \textit{i.e.} all charges which can be determined
in $V$ have sharp values within a given charge class. We also 
note  that the restricted representations can be
reconstructed from the partial states. Thus only data 
which can be taken in $V$ are needed in order to fix the charge classes.
We therefore propose to replace the notion of superselection sector
by the (in the presence of massless particles
more realistic) concept of charge class. 

\section{Charged morphisms}

In order to clarify the structure of the family of charge classes one 
has to understand their mutual relation. Given $V$, one can
proceed from the partial states in the charge class of the 
vacuum to the partial states in a given charge class by limits of local 
operations in $V$. Heuristically, these operations may be 
thought of as creation of pairs of opposite charges on some 
given time shell, where the unwanted compensating charge is 
shifted to lightlike infinity; it thereby disappears in the
spacelike complement of any relatively compact region in $V$.
In order to control the energy required for these operations
one has to localize them in broadening 
hypercones $\cL$, cf.\ Fig.~3. (A hypercone is
the causal completion of an open pointed convex cone 
formed by geodesics on some time shell in $V$.) 
These heuristic considerations are 
put into mathematically precise form as follows.

\begin{figure}[h]
\centering 
\epsfig{file=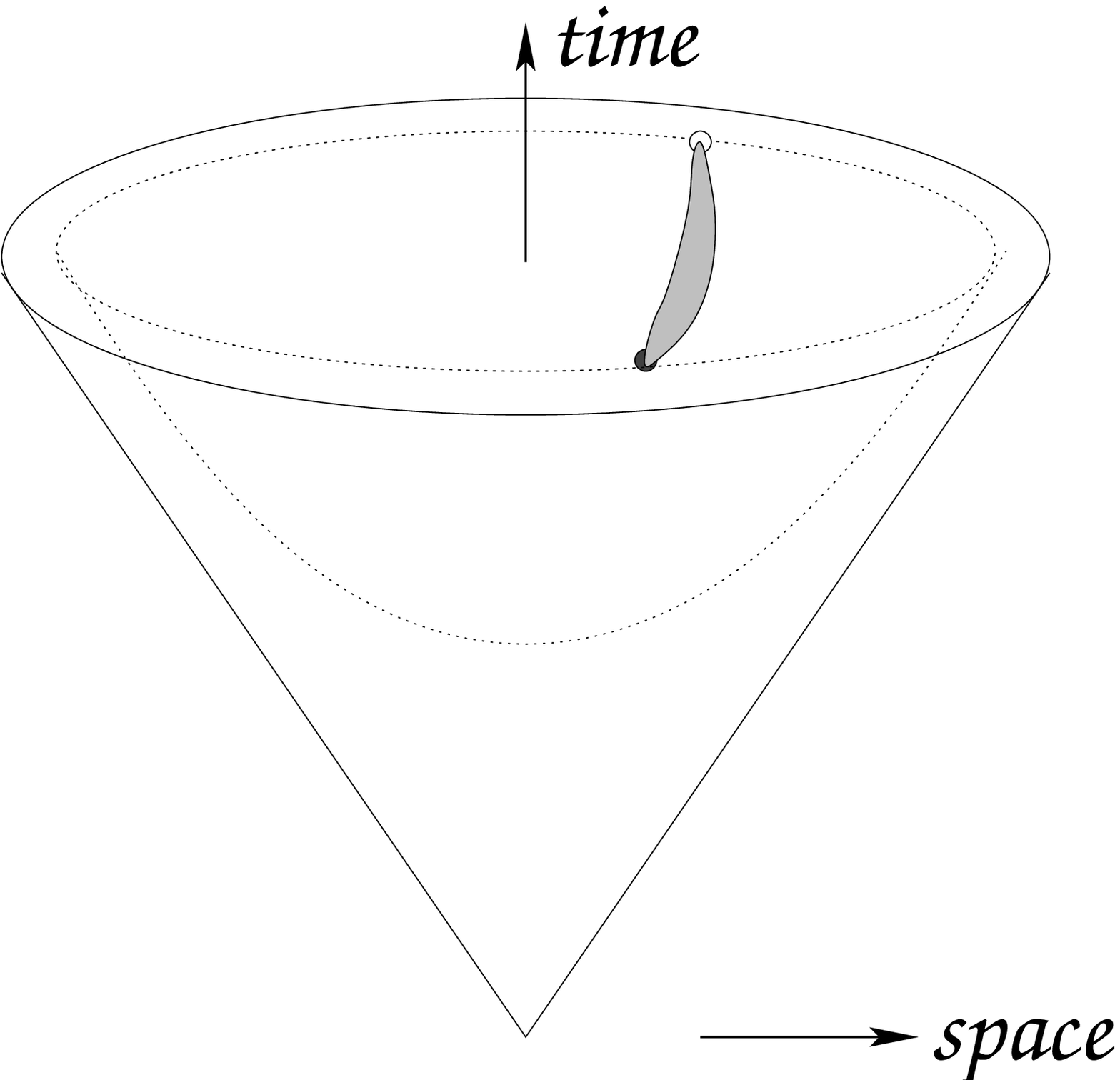,width=55mm} 
 \hspace*{12mm}
\epsfig{file=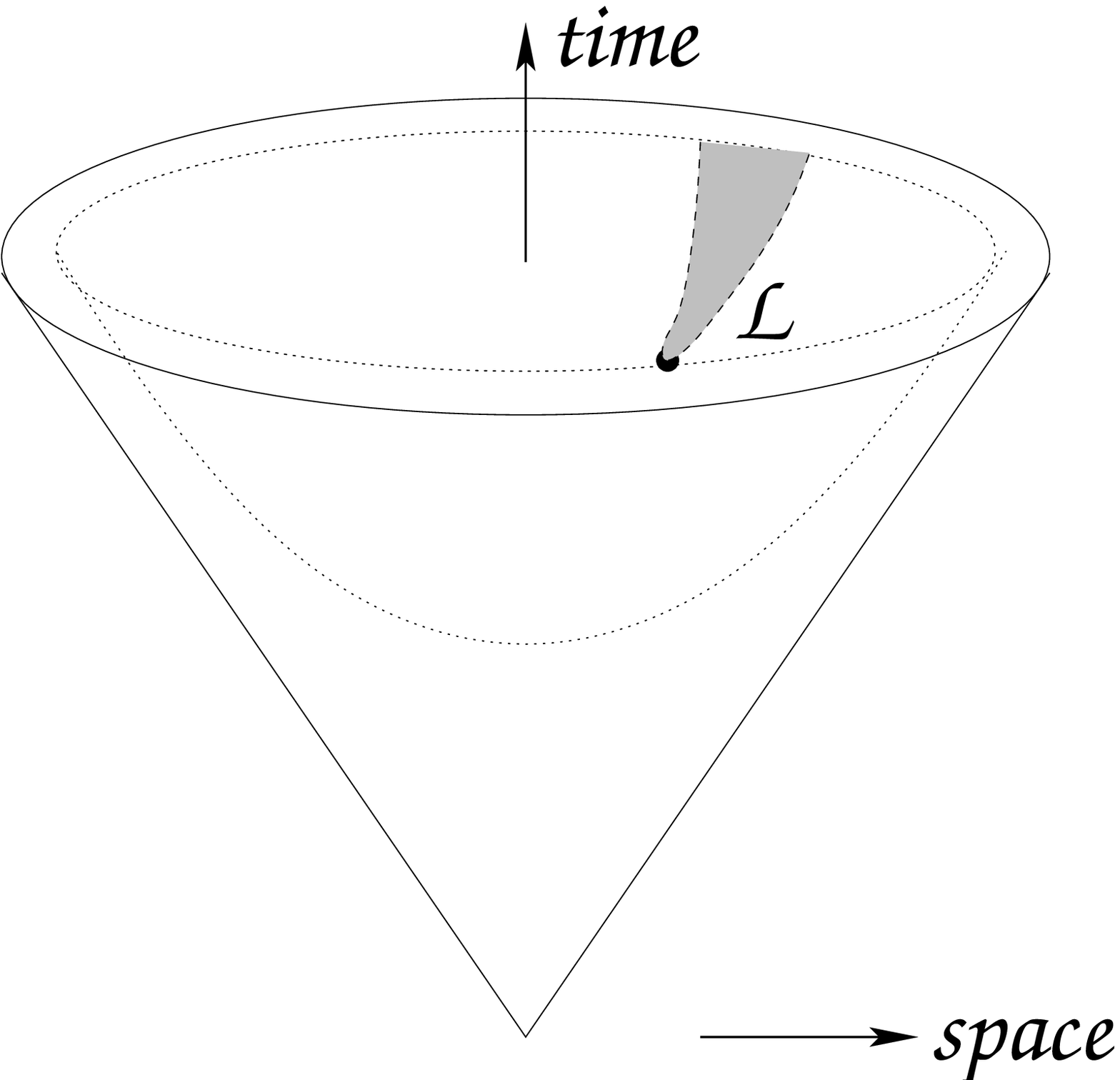,width=55mm} 
\caption{Local operation creating a pair 
and charge creation in a hypercone $\cL$ as a limit case}             
\label{fig3}
\epsfxsize20mm
\end{figure}

\noindent {\bf Assumption:} Given a charge class, there exists for any 
hypercone $\cL \subset V$ a sequence of inner automorphisms 
$\{ \sigma_n \in \text{\rm In} \, {\cA}(\cL) \}_{n \in \NN}$, 
induced by unitaries in ${\cA}(\cL)$, such that                           
$$  \rho_\cL \doteq \mbox{w}-\lim_{n} \sigma_n \quad 
\mbox{exists pointwise on}  
\ \ \cA(V)$$ 
and the adjoint of $\rho_\cL$ maps the partial states in the 
charge class of the vacuum onto the partial states in the 
target charge class. 

\vspace*{2mm} \noindent 
The properties of the limit maps are summarized in the following proposition.

\begin{proposition} For fixed charge class and any $\cL \subset V$, 
let $\rho_\cL : \cA(V) \rightarrow \cA(V)^-$ 
be the map defined above. 

\vspace*{-2mm}
\begin{enumerate}
\item[(a)] \ $\rho_\cL$ is linear,  symmetric  and multiplicative          
\item[(b)] \ $\rho_\cL \upharpoonright {\cA}(\cR) = \iota$ \ if \
$\cR \mbox{\Large $\times$} \cL$
\item[(c)] \ $ \rho_\cL \, ({\cA}(\cR))^- \subseteq
{\cA}(\cR)^{-}$ \ if \ $\cR \supseteq {\cal L}$. 
\item[(d)] \ $\rho_{\cL_1} \simeq \rho_{\cL_2}$ for any pair 
of hypercones  $\cL_1, \cL_2 \subset V$. 
\end{enumerate}
\end{proposition}

According to point (a) these 
maps define representations of the                            
algebra $\cA(V)$ for the given lightcone $V$. 
Points (b) and (c) encode the information that they 
arise from  local operations in $\cL$ and point (d)
expresses the fact that infrared clouds, which                    
are inevitably produced by the charge creating operations, 
cannot sharply be discriminated in $V$. In analogy
to the terminology used in sector analysis, 
these maps are called (hypercone) localized morphisms.

We restrict attention here to the simplest, but physically 
important family of charge classes where in part (c) of 
the proposition equality holds for the corresponding 
morphisms. Then 
$\rho_\cL(\cA(V))^- = \cA(V)^-$ is a factor of 
type III${}_1$. Taking the fact into account that the space
of normal states on such factors is homogeneous \cite{CoSt}, 
\textit{i.e.} the (adjoint) inner automorphisms of 
$\cA(V)^-$ act almost transitively on normal states, 
it is meaningful to assume that there exist morphisms
as in point (d) of the proposition which are 
related by unitary intertwiners in $\cA(V)^-$. We summarize 
these features in the following definition. \\[2mm]              
{\bf Definition of simple charge classes:} \ A charge class is 
said to be simple if for any hypercone $\cL \subset V$ there
exist corresponding localized morphisms $\rho_\cL$ such that 
\begin{enumerate}
\item[(i)] $ \rho_\cL \, ({\cA}(\cR))^- = {\cA}(\cR)^{-}$ if 
$\cR \supseteq {\cal L}$ 
\item[(ii)] $\rho_{\cL_1} \simeq \rho_{\cL_2}$ for any pair 
of hypercones  $\cL_1, \cL_2 \subset V$ and there exist 
corresponding unitary intertwiners in $\cA(V)^-$.    
\end{enumerate} 
Localized morphisms attached to a simple charge class
are said to be simple. 

\section{Statistics and symmetries}

Similarly to the case of superselection sectors, one has to rely in the 
analysis of charge classes on a maximality condition for the 
hypercone algebras, akin to Haag duality:  
for any hypercone $\cL \subset V$ one has
$$\cA(\cL)^\prime \cap \cA(V)^- = \cA(\cL^c)^- \, , \quad 
\cA(\cL^c)^\prime \cap \cA(V)^- = \cA(\cL)^- \, , $$
where $\cL^c$ denotes the spacelike complement of $\cL$ in $V$.
Substantial results supporting this form of hypercone duality have
been established in \cite{Ca} for the free Maxwell field. 

It is an important consequence of hypercone duality that
equivalent morphisms which are localized in
neighboring hypercones $\cL_1, \cL_2$ have intertwiners which 
are contained in $\cA(\cL)^-$, where $\cL$ is any larger hypercone
containing $\cL_1$ and $\cL_2$. Making use of this fact one can extend 
the morphisms from their domain $\cA(V)$ to larger algebras containing
the weak closures of certain hypercone algebras. Based on these
extensions, the following result for the 
family of simple localized morphisms has been established.

\begin{proposition} \label{5.1}
Let $\rho_1$, $\rho_2$ be simple morphisms which are 
localized in hypercones $\cL_1, \cL_2$, respectively.
\begin{enumerate}
\item[(a)] The (suitably extended) morphisms can be composed and 
there is for any given hypercone $\cL$ some simple morphism $\rho$ localized
in $\cL$ such that $\rho_1 \circ \rho_2 \simeq \rho$. 
\item[(b)] $\rho_1 \circ \rho_2 \simeq  \rho_2 \circ \rho_1$. \  
If $\rho_1$, $\rho_2$ belong to the same charge class there is some
canonical intertwiner $\varepsilon(\rho_1,\rho_2) \in \cA(V)^-$
depending only on the given morphisms. 

\item[(c)] For each charge class there exists a statistics parameter 
$\varepsilon \in \{ \pm 1 \}$ such that for any pair of 
morphisms $\rho_1$, $\rho_2$ in  this class which are localized
in spacelike separated hypercones $\cL_1, \cL_2$, cf.\ Fig.~4, 
one has  \ $\varepsilon({\rho_1, \rho_2}) = \varepsilon \, 1$.  

\vspace*{5mm}
\begin{figure}[h]
\centering 
\epsfig{file=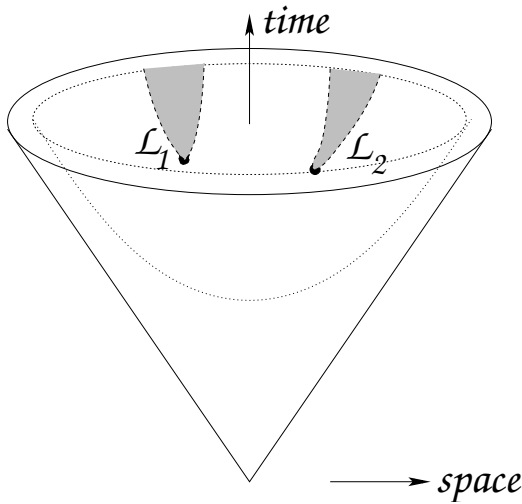,width=70mm}
\caption{Spacelike separated hypercones}                            
\label{fig5}
\end{figure}

\item[(d)] For each simple charge class there exists a simple conjugate
charge class such that for any morphism $\rho$ in the given class
there is a corresponding morphism $\overline{\rho}$ in the conjugate
class satisfying  
$\rho \circ \overline{\rho} = \overline{\rho} \circ \rho =
\iota$. Moreover, the conjugate class has the same statistics
parameter as the given class. 
\end{enumerate}
\end{proposition}
These results do not depend on the choice of lightcone $V$.
According to item (c) any simple charge class has definite  
(Bose respectively Fermi) statistics  
and item~(d) says that to each simple charge class there is a 
simple conjugate class of states carrying opposite (neutralizing)
charges. Moreover, items (a), (b) and 
(d) imply that the equivalence classes of simple morphisms 
determine an abelian group with the product given by composition. Its     
dual can be interpreted as global gauge group generated 
by the simple charges which can be sharply determined in
lightcones.                                                          
Thus the simple charge classes have a 
structure similar to that of the simple sectors in massive theories.
The elusive theoretical effects of the infrared clouds 
completely disappear from the discussion by taking into proper 
account the spacetime limitations of real experiments.             

\section{Covariance and spectrum}

In the discussion of the covariance properties of simple charge 
classes and of their energetic features,  one has to take 
into account that one has  merely an                                  
endomorphic action of the semigroup 
${\cal S}^\uparrow_+ \doteq \overline{V}_+ 
\rtimes {\cal L}^\uparrow_+ \subset {\cal P}^\uparrow_+$
on any lightcone  $V$. 
The following characterization of covariant simple
morphisms is appropriate in this case. \\[2mm]
{\bf Definition of covariant morphisms:} \ 
Let $\rho : {\mathfrak A}(V) \rightarrow 
{\mathfrak A}(V)^-$ be a simple 
morphism which is localized in $\cL$. The morphism is said to
be covariant if it is the initial member of
a family of 
morphisms $\{ {}^\lambda \! \rho \}_{\lambda \in {\cal
  S}^\uparrow_+} $ which, for given $\lambda \in  {\cal
S}^\uparrow_+$, are localized 
in any hypercone $ {}^\lambda \cL \supset \lambda \cL$, satisfy
the relation 
$$   {}^{\lambda \mu} \! \rho \circ \alpha_\lambda =  
\alpha_\lambda \circ  {}^{\mu} \! \rho \, , \quad 
\lambda, \mu \in  {\cal S}^\uparrow_+ $$
and are affiliated with the same charge 
class as $\rho \equiv {}^1 \! \rho$.
More precisely, there exists a weakly continuous section of       
intertwiners $\lambda \mapsto \Gamma_\lambda \in \cA(V)^-$
between ${}^\lambda \! \rho$ and $\rho$. \\[2mm] 
It can be shown that the family of morphisms affiliated in this 
manner with a given morphism is unique.
The properties  of covariant morphisms
are described in the following proposition. 

\begin{proposition} Consider the family of covariant simple morphisms
which are localized in hypercones contained in a given lightcone $V$.      
\begin{itemize}
\item[(a)] The family is stable under composition and conjugation
and all results of Proposition~\ref{5.1} apply to it. 
\item[(b)] Each morphism $\rho$
determines a unique unitary representation $U_\rho$ of  
(the covering of) the full Poincar\'e
group $\widetilde{\cal P}^\uparrow_+ = \RR^4 \rtimes \widetilde{\cal
  L}^\uparrow_+$  such that  
$$ \mbox{Ad} \, U_\rho(\tilde{\lambda}) \circ \rho 
= \rho \circ \alpha_\lambda \, , \quad
\widetilde{\lambda} \in \widetilde{\cal S}^\uparrow_+ $$
\item[(c)] $\mbox{sp} \, U_\rho \upharpoonright \RR^4 \subset \overline{V}_+$
\item[(d)] In presence of massless
particles \ $U_\rho(\RR^4) \, \mbox{\boldmath $\notin$} \, \cA(V)^-$.
\end{itemize}
\end{proposition}

This result shows that the covariant morphisms
describe the expected physical properties of
elementary systems in a meaningful way. In particular, it is 
possible to assign with the help of the unitary groups  
established in (b) an energy content to the partial 
states on $\cA(V)$ within a given charge class. This energy 
is bounded from below according to (c),  
expressing the stability of the states. But in view of 
point~(d) the generators of the time 
translations should not be interpreted as genuine observables.
This can be understood if one bears in mind 
that part of the energy  content of a global 
state may be lost by outgoing radiation created in the 
past of~$V$. Phrased differently, the energy content
of the partial states on $\cA(V)$
is fluctuating. Thus the generators of
the time translations in (b) require an interpretation
similar to that of the Liouvillians in quantum 
statistical mechanics.

\section{Concluding remarks}

In this work the origin of the infrared difficulties in the 
interpretation of theories with long range forces has been traced
back to the unreasonable idealization of observations covering
all of Minkowski space. Observations are at best performed in 
future directed lightcones, hence the arrow of time enters already 
in the interpretation of the microscopic theory.

As was explained,
the restriction of states to the observables in any given 
lightcone $V$ amounts to a geometric, Lorentz invariant                   
infrared regularization. It allows to form charge
classes of states carrying the same global charges but 
differing in their infrared features. The pertinent observable 
algebras ${\cA}({V})$ are highly reducible due to the loss of
information about outgoing radiation created in the past of $V$. Yet the data 
which can be obtained in $V$ are sufficient to determine
sharply the global charges, their statistics and the underlying   
gauge group. These results have been established so far only for 
simple charge classes; but work in progress indicates that
they hold more generally.

The data obtainable in lightcones $V$ are also sufficient to fix for each
charge class a representation $U_V$ of the Poincar\'e group,
respectively of its covering. Yet the generators of this representation
cannot be interpreted as genuine observables since they 
incorporate in a gross manner fluctuations of the 
background radiation. In this respect the situation resembles 
the treatment of ensembles in quantum statistical mechanics.

Since infrared sectors cannot be discriminated in any lightcone
$V$ it seems likely that also the infraparticle problem 
(failure of the Wigner concept of particle for electrically
charged states \cite{FrMoSt,Bu2}) 
disappears if one resorts to the representations $U_V$ of the Poinar\'e group. 
In fact, the radiation observed in $V$ can be described by 
Fock states with a finite  particle number; so there may well exist
partial states on $\cA(V)$, describing a single electrically 
charged particle where the (globally inevitable) accompanying
radiation field has no observable effects in $V$. Such states
could well contribute to an atomic part in the mass spectrum of
$U_V$.

\end{document}